\documentclass[letterpaper, 10 pt, conference]{ieeeconf}
\usepackage{cite}
\usepackage{hyperref}
\usepackage{color}
\usepackage{listings}
\usepackage{float}
\usepackage{orcidlink}
\usepackage{glossaries}

\newacronym{EHR}{EHR}{electronic health record}
\newacronym{DSL}{DSL}{domain-specific language}
\newacronym{HTTP}{HTTP}{Hypertext Transfer Protocol}
\newacronym{REST}{REST}{representational state transfer}

\glsunset{HTTP}
\glsunset{REST}

\glsdisablehyper

\begin{document}

\title{\LARGE \bf
    The Path to a Modular and Standards-based Digital Health Ecosystem
}

\author{
    Paul Schmiedmayer \orcidlink{0000-0002-8607-9148}, 
    Vishnu Ravi \orcidlink{0000-0003-0359-1275}, 
    Oliver Aalami \orcidlink{0000-0002-7799-2429}
    \\
    \textit{Byers Center for Biodesign, Stanford University}, Stanford, CA, USA
}

\maketitle
\thispagestyle{empty}
\pagestyle{empty}

\begin{abstract}
Software engineering for digital health applications entails several challenges, including heterogeneous data acquisition, data standardization, software reuse, security, and privacy considerations.
We explore these challenges and how our Stanford Spezi ecosystem addresses these challenges by providing a modular and standards-based open-source digital health ecosystem. 
Spezi enables developers to select and integrate modules according to their needs and facilitates an open-source community to democratize access to building digital health innovations.
\newline

\indent \textit{Clinical relevance}— 
The Stanford Spezi ecosystem addresses crucial challenges in developing digital health applications, including interoperability, privacy, security, and scalability, enabling new approaches and research in prevention, observation, and patient care built on our experience of supporting over 20 projects over several years. 
\end{abstract}

\section{Introduction}
\label{sec:Introduction}

Modern digital health solutions facilitate longitudinal data collection and analysis, yielding valuable insights for patients and healthcare professionals.
They offer the potential for continuous patient monitoring, enabling preventative interventions and cost-effective home-based care.
Developing such software demands a deep understanding of medical applications, interoperability and privacy standards, and software engineering principles.
Therefore, creating reusable, scalable, and evolvable systems is essential.

After successfully designing and deploying our first digital health application, VascTrac, in 2016~\cite{ata2018vasctrac} and recognizing the hurdles in transitioning from idea to prototype, we established the open-source CardinalKit template application in 2018~\cite{aalami2023cardinalkit}.
The template application served as an initial guide for developers creating mobile digital health applications, including students in our \textit{Building for Digital Health} class.
Over the years, we have enriched the open-source template with additional functionality and integration with the HIPAA-compliant Stanford mHealth platform.

\section{Challenges}
\label{sec:Challenges}

While the initial development of the CardinalKit template provided a valuable foundation for digital health projects, we observed significant effort being spent on customizing the template to fit individual project needs.
Additionally, adopting modern data standards such as HL7 FHIR was not baked into the architecture, and the framework's flexibility, especially concerning architecture, cloud services, and authentication, needed to be improved.

Maintaining and expanding this framework posed its own set of challenges, prompting a reevaluation of our strategy.
We aimed to enhance selected reusability and accessibility for newcomers while preserving its functional and developer-friendly nature.
Moreover, we sought to foster a more robust open-source community around our software, empowering individuals to take charge of components and promoting collaborative efforts.
\section{Solution}
\label{sec:Solutions}

To address these challenges, we completely revised our development approach and rewrote our open-source software from the ground up, to reconceptualize a digital health development ecosystem.
The \textbf{Stanford Spezi}\cite{Schmiedmayer_Spezi} ecosystem emerged as a modular, standards-based digital health system, enhancing the template application-based approach with a diverse software package ecosystem.

Rather than pre-packaging software components, we devised independent modules that interact through standardized methods.
This allows developers to progressively adopt varying functionalities, akin to building with Lego pieces, and cater to their specific needs.
With a foundation in standards-based data exchange like HL7 FHIR, developers can reuse components and connect mobile, wearable, and spatial digital health solutions to \gls{EHR} systems.

We urge developers and project partners to create, control, and upkeep their modules.
We envision an indexed collection of discoverable that meet stringent quality standards.
Offering smaller components aims to make contribution and innovation more accessible and facilitate quicker iterations.

Several modules have been instantiated, encompassing critical aspects of the earlier template application. 
We plan to broaden the architecture to include the Google Android platform, develop additional dashboard and \gls{EHR} integration components, and create modules based on project partnerships' needs.

\section*{Acknowledgment}
We want to thank our project partners, open-source contributors, and especially Ashley Griffin, Santiago Gutierrez, and Varun Shenoy for their CardinalKit code contributions.

\bibliographystyle{IEEEtran}
\bibliography{bibliography}

\begin{thebibliography}{1}
\providecommand{\url}[1]{#1}
\csname url@samestyle\endcsname
\providecommand{\newblock}{\relax}
\providecommand{\bibinfo}[2]{#2}
\providecommand{\BIBentrySTDinterwordspacing}{\spaceskip=0pt\relax}
\providecommand{\BIBentryALTinterwordstretchfactor}{4}
\providecommand{\BIBentryALTinterwordspacing}{\spaceskip=\fontdimen2\font plus
\BIBentryALTinterwordstretchfactor\fontdimen3\font minus
  \fontdimen4\font\relax}
\providecommand{\BIBforeignlanguage}[2]{{%
\expandafter\ifx\csname l@#1\endcsname\relax
\typeout{** WARNING: IEEEtran.bst: No hyphenation pattern has been}%
\typeout{** loaded for the language `#1'. Using the pattern for}%
\typeout{** the default language instead.}%
\else
\language=\csname l@#1\endcsname
\fi
#2}}
\providecommand{\BIBdecl}{\relax}
\BIBdecl

\bibitem{ata2018vasctrac}
\BIBentryALTinterwordspacing
R.~Ata, N.~Gandhi, H.~Rasmussen, O.~El-Gabalawy, S.~Gutierrez, A.~Ahmad,
  S.~Suresh, R.~Ravi, K.~Rothenberg, and O.~Aalami, ``Clinical validation of
  smartphone-based activity tracking in peripheral artery disease patients,''
  \emph{npj Digital Medicine}, vol.~1, no.~1, p.~66, Dec 2018. [Online].
  Available: \url{https://doi.org/10.1038/s41746-018-0073-x}
\BIBentrySTDinterwordspacing

\bibitem{aalami2023cardinalkit}
\BIBentryALTinterwordspacing
O.~Aalami, M.~Hittle, V.~Ravi, A.~Griffin, P.~Schmiedmayer, V.~Shenoy,
  S.~Gutierrez, and R.~Venook, ``Cardinalkit: Open-source standards-based,
  interoperable mobile development platform to help translate the promise of
  digital health,'' \emph{JAMIA Open}, 2023. [Online]. Available:
  \url{https://doi.org/10.1093/jamiaopen/ooad044}
\BIBentrySTDinterwordspacing

\bibitem{Schmiedmayer_Spezi}
\BIBentryALTinterwordspacing
P.~Schmiedmayer, V.~Ravi, and O.~Aalami, ``{Spezi}.'' [Online]. Available:
  \url{https://doi.org/10.5281/zenodo.8013043}
\BIBentrySTDinterwordspacing

\end{thebibliography}

\end{document}